# Preprint - Controlled Transition Metal Nucleated Growth of Carbon Nanotubes by Molten Electrolysis of CO$_2$

Xinye Liu [1], Gad Licht [2], Xirui Wang [1] and Stuart Licht [1,2,*]

[1] .Department of Chemistry, George Washington University, Washington DC 20052, USA; slicht@gwu.edu (SL)
[2] C2CNT, 1035 26 St NE, Calgary, AB T2A 6K8, Canada; sl@c2cnt.com (SL)
* Correspondence: slicht@gwu.edu

**Abstract:** The electrolysis of CO$_2$ in molten carbonate has been introduced as an alternative mechanism to synthesize carbon nanomaterials inexpensively at high yield. Until recently, CO$_2$ was thought to be unreactive, making its removal a challenge. CO$_2$ is the main cause of anthropogenic global warming and its utilization and transformation into a stable, valuable material provides an incentivized pathway to mitigate climate change. This study focuses on controlled electrochemical conditions in molten lithium carbonate to split CO$_2$ absorbed from the atmosphere into into carbon nanotubes, and into various macroscopic assemblies of CNTs,, which may be useful for nano-filtration. Different CNTs, morphologies were prepared electrochemically by variation of the anode and cathode composition and architecture, electrolyte composition pre-electrolysis processing, and the variation of current application and current density. Individual CNT morphologies structures and the CNT molten carbonate growth mechanism are explored by SEM, TEM, HAADF EDX, XRD and Raman. The principle commercial technology for CNT production had been chemical vapor deposition, which is an order of magnitude more expensive, generally requires metallo-organics, rather than CO$_2$ as reactants, and can be highly energy and CO$_2$ emission intensive (carries a high carbon positive, rather than negative, footprint).

## 1. Introduction

Global CO$_2$ has risen rapidly accelerating extinction risk [1-4]. CO$_2$ is a highly stable molecule and difficult to remove from the environment [5]. One means to mitigate CO$_2$ under consideration is its low energy chemical transformation to a (1) stable, (ii) useful (iii) valuable product, with a low cost and low carbon footprint of production. The transformed CO$_2$'s product stability prevents the captured CO$_2$ from re-emission, product usefulness provides a buffer to store the capture carbon, and high-value (ideally higher than the cost of CO$_2$ transformation) provides an economic incentive to remove the greenhouse gas. Graphitic nanocarbons, such as carbon nanotubes made from CO$_2$, may meet several of these transformed CO$_2$ product requirements. For example, its basic structure of layered graphene retains the durability of graphite, whose hundreds of millions year old mineral deposits attests to its long term stability, while CNT's market value of $100,000 to $400,000 / tonne can provide a revenue, rather than a cost, while removing CO$_2$.

Multiwalled CNTs (**C**arbon **N**ano**T**ubes) are comprised of concentric cylindrical graphene sheets. CNTs have a measured tensile strength of 93,900 MPa, which is the highest tensile strength of any material [6,7]. Other useful properties of CNTs include high electrical, high thermal conductivity, flexibility, high capacity for charge storage, and catalysis. CNTs applications range from stronger, lighter structural materials including cement, aluminum, and steel admixtures [8], medical applications [9,10], elec-tronics, batteries and supercapacitors [11,12], sensors and analysis [13-15], plastics and polymers [16-20] textiles [21], hydrogen storage [22] and water treatment [23,24].

The deliberate thought of this study is that the superior physical chemical properties of CNTs, and in particular CNTs made by consuming $CO_2$, will cause a demand for its application incentivizing $CO_2$ consumption and driving climate change mitigation by decreasing emissions of the greenhouse gas $CO_2$.

Increased pathways for the use of $CO_2$ as a molten carbonate electrolysis reactant to synthesize value-added CNTs will provide the effect of opening a path to consume this greenhouse gas to mitigate climate change, and its transformation to CNTs will provide a stable material to store carbon removed from the environment.

To date the CNT' market has been limited due to a high cost of production. Commercially, CNTs are mainly produced by chemical vapor deposition, CVD and not from $CO_2$ [25,26]. CVD production of CNTs is chemical and energy intensive and expensive, leading to current costs of $100K to $400K per tonne CNT, and CVD production has a high carbon footprint [27].

Prior attempts to transform $CO_2$ to carbon nanotubes or graphene have been low yield and energy intensive such as the production of graphitic flakes using high pressure $CO_2$ or by electrolysis in molten $CaCl_2$ electrolytes. Undesired byproducts included $Al_2O_3$, hydrogen, and hydrocarbons, and from the electrolysis, carbon monoxide byproducts from an 850°C electrolysis splitting in molten $CaCl_2$ electrolytes [28,29].

However, $CO_2$ has a strong affinity for certain molten carbonates. In 2009 a process was introduced to mitigate the greenhouse gas $CO_2$ through molten electrolytic splitting and transformation at elevated temperatures [31]. Pathways were opened to the high purity, renewable energy electrolytic splitting of $CO_2$ to solid carbon by demonstrating that in molten lithium carbonate (melting point 723 °C) electrolytes, the 4 electron molten electrolysis reduction of tetravalent carbon to solid products dominates, below 800 °C. With rising electrolysis temperature between 800 and 900°C the 2 electron reduction to a CO product increasingly dominates, and by 950°C the transition to the alternative CO byproduct is complete [30].

The solid carbon product of $CO_2$ electrolysis was further refined to graphitic nanocarbons through the discovery of catalyzed molten electrolysis transition metal nucleated growth of carbon nanotubes and carbon nanofibers in lithium carbonate electrolytes [31-33]. The process was given the acronym C2CNT (<u>C</u>arbon dioxide to <u>C</u>arbon <u>N</u>ano<u>T</u>ubes). The $^{13}C$ isotope of $CO_2$ was used to track carbon through the C2CNT process from its origin ($CO_2$ as a gas phase reactant) through its transformation to a CNT or carbon nanofiber product, and that the $CO_2$ originating from the gas phase serves as the renewable C building blocks in the observed CNT product [32]. The net reaction is:

Dissolution: $CO_2(gas) + Li_2O(soluble) \rightleftharpoons Li_2CO_3(molten)$ (1)
Electrolysis: $Li_2CO_3(molten) \rightarrow C(CNT) + Li_2O (soluble) + O_2(gas)$ (2)
Net: $CO_2(gas) \rightarrow C(CNT) + O_2(gas)$ (3)

The electrolytic $CO_2$ splitting in molten carbonates can occur at electrolysis potentials less than 1 volt [33]. Due to the high affinity of $CO_2$ towards reaction 1, that is for the $Li_2O$ present in the electrolyte, in the C2CNT process, the electrolytic splitting can occur as direct air carbon capture air without $CO_2$ pre-concentration, or with exhaust gas, or with concentrated $CO_2$. As it directly captures $CO_2$ from the air, no further introduction of $CO_2$ is needed throughout the group's experiments. By variation of the electrolysis setup the process can produce in addition to conventional morphology CNT: doped CNTs, helical CNTs and magnetic CNTs as illustrated in Figure 1 [31-44]. Studied applications of electrolytic CNTs from $CO_2$ include batteries [34], $CO_2$ transformation from power station flue gas [45] and the substantial decrease in the carbon footprint of structural materials as CNT-composites including: CNT-cement, CNT-steel and CNT-aluminum [8,46] as well as modification of the $CO_2$ splitting process to yield other CNMs including carbon nano-onions, carbon platelets and graphene [47-51].

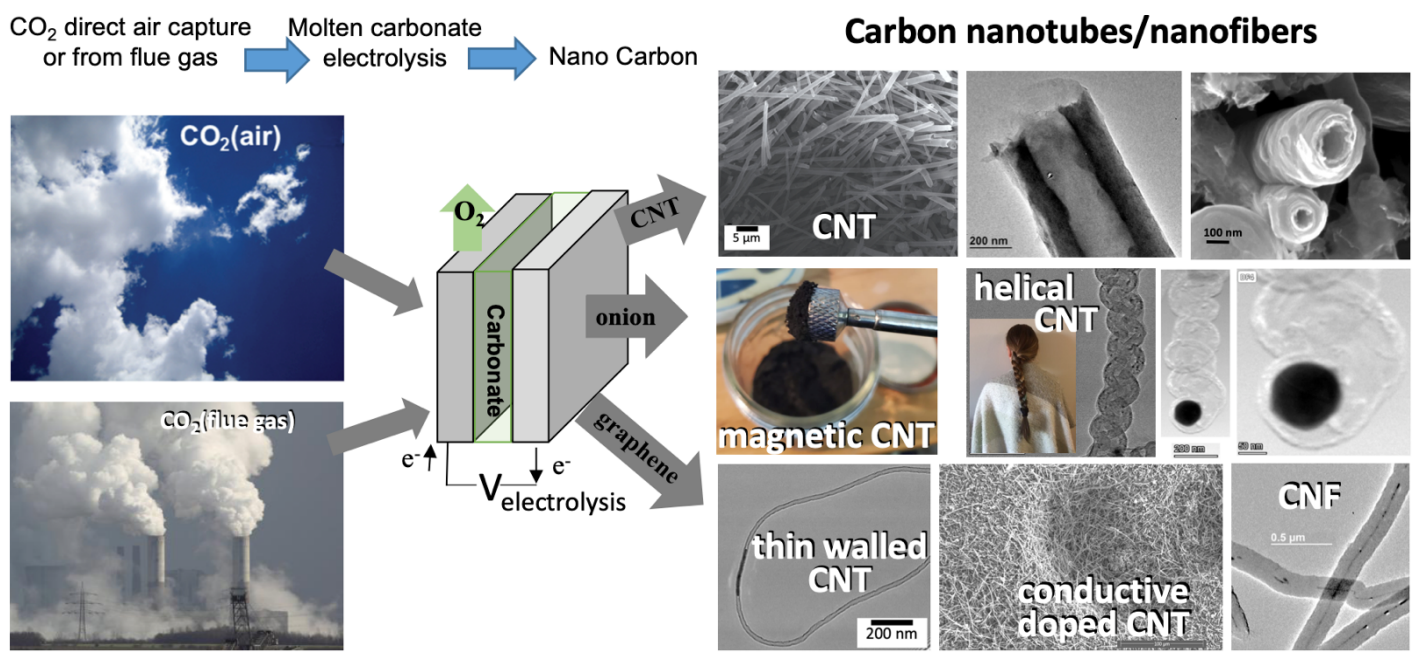

**Figure 1.** High yield electrolytic synthesis of carbon nanotubes from $CO_2$, either directly from the air or from smokestack $CO_2$, in molten carbonate.

The rise of this greenhouse gas is causing extensive climate change and damage to the planet's ecosphere, and its mitigation is one of the most pressing challenges of our time [1-5]. Technical, catalyst driven solutions to mitigate climate change are of the highest significance to the catalyst, not only due to their probes of a new chemistry to catalyze nanocarbon formation, but also by galvanizing the community with action towards mitigation of the existential climate change threat facing the planet. This study provides four contributions to a catalyst driven solution to climate change (i) 10 distinct, new electrochemical procedures are presented to transform $CO_2$ to CNTs at high purity. The procedures produce a variety of distinct CNT morphologies ranging from curled to straight, short to long, and thin to thick. (ii) This study explores the transition metal nucleation that catalyzes the process to produce high purity carbon nanotubes. (iii) The study demonstrates new syntheses of macroscopic assemblies of CNTs by the C2CNT process, with structural implications towards their potential applications for nano-filtration and neural nets. (iv) This study provides an extensive carbon nanotube baseline to a companion study in which the same electrochemical components are utilized in new configurations to generate entire new classes of non-carbon nanotube graphitic nanocarbons.

## 2. Results and Discussion

*2.1. Electrolytic conditions to synthesize high purity, high yield CNTs from $CO_2$.*

The first part of this study systematically explores electrochemical parameters to reveal a wide variety of conditions that yield a high purity, high yield CNT product by electrolysis of $CO_2$ in 770°C lithium carbonate. An in depth look at the material composition and morphologies of the products is conducted, particularly around the transition metal nucleation zone of CNT growth. The latter part of this study reveals molten electrochemical conditions which produce macroscopic assemblies of CNTs. This study also serves as a sister study [51] in which small electrolytic changes in the 770°C molten

Li$_2$CO$_3$ yield major changes to the product consisting of new, non-CNT nanocarbon allotropes).

Previously, we had shown that the high production rate (using a high electrolysis current density, J, of 0.6 A/cm$^2$) electrolytic splitting of CO$_2$ in molten Li$_2$CO$_3$ electrolyte using a Muntz Brass cathode (60% Cu and 40% Zn) and a Nichrome C (60% Ni, 24% Fe and 16% Cr) anode produced high quality (97% purity), high aspect ratio carbon nanotube (CNT) product with addition to the electrolyte of either 0.1 wt% Fe$_2$O$_3$ [43] or 2.0 wt% Li$_2$O [41,44]. Addition of higher concentrations of either iron or lithium oxide to the electrolyte increased the formation of defects in the CNTs, as measured by Raman spectroscopy, which at this higher current density induced spiraling of the CNT during growth and the observation of the controlled growth of a variety of helical carbon nano-allotropes including single and double braided helices as well as flat, spiral morphologies [43,44].

Here, conditions related to the high purity CNT synthesis are systematically varied to determine other electrochemical conditions which support the high purity, low defect formation of straight (non-helical) CNTs. Examples of the conditions which are varied are composition of the cathode and anode, additives to the lithium carbonate electrolyte and current density and time of the electrolysis. Variations of the electrodes include the use of cathode metal electrodes such as Muntz brass Monel, or Nichrome alloys. Anode variations include noble anodes such as iridium, various nickel containing anodes including nickel, Nichrome A or C, Inconel 600, 625, or 718, or specific layered combinations of these metals. Electrolyte additives that are varied include Fe$_2$O$_3$, and nickel or chromium powder, and electrolyses are varied over a wide range of electrolysis current densities. Several electrolyses studied here which yield high purity, high yield carbon nanotubes are described in Table 1. Scanning electron microscopy (SEM) of the products of a variety of those CNT syntheses as conducted by CO$_2$ electrolysis in molten Li$_2$CO$_3$ at 770°C are presented in Figure 2.

For Electrolysis #A, the top row of Table 1 presents electrochemical conditions, and the row of Figure 2 presents SEM of product, of a repeat of the electrochemical conditions of the described 58 0.1 wt% Fe$_2$O$_3$ electrolysis (same lithium carbonate electrolyte, same Muntz Brass cathode and Nichrome C anode, same 0.6 A/cm$^2$ current density and 30 minute electrolysis duration), but uses a simpler (from a material perspective) alumina (ceramic Al$_2$O$_3$), rather than stainless steel 304, electrolysis cell casing. Use of the alumina casing in this study limits the pathways for metals to enter reducing parameters to evaluate, and possibly effect, the electrolytic system. Note however, that the stainless steel 304 had not been observed to corrode, and the switch from stainless to alumina was not observed to materially affect the electrolysis product. The CNT product is again 97% purity, coulombic efficiency is 99%, which quantifies the measured available charge (current multiplied by the electrolysis time) to the measured number of 4 electrons per equivalent of C in the product, and the carbon nanotube length is 50 to 100 μm.

The second row of Figure 2 (panels #B) changes only the current density, which is lowered to 0.15 A/cm$^2$ and the electrolysis time, which is increased to 4 hours, and the result is a decrease in product purity to 94%, a decrease in CNT length to 20-80 μm, and a modest decrease in coulombic efficiency to 98%. At this current density, as observed in the third row of Figure 2, panels #C, addition of 0.1 wt% Ni along with the 0.1 wt% Fe$_2$O$_3$, results in 96% purity zigzag, twisted, rather than straight CNTs. These twists can be induced by over-nucleation decreasing control of the CNT linear growth. In the most magnified of these product images (right side of the Figure, 2 μm bar resolution) evidence of the over-nucleation is observed in the larger nodules visible at the CNT tips and joints.

**Table 1.** A variety of Electrolytic CO$_2$ splitting conditions in 770°C Li$_2$CO$_3$ producing a high yield of carbon nanotubes.

| Electrolysis # | Cathode | Anode | Additives (wt% powder) | Electr time | Current density A/cm² | Product Description |
|---|---|---|---|---|---|---|
| **A** | Muntz Brass | Nichrome C | 0.1%Fe$_2$O$_3$ | 0.5h | 0.6 | 97% Straight 50-100 μm CNT |
| B | Muntz Brass | Nichrome A | 0.1%Fe$_2$O$_3$ | 4h | 0.15 | 94% Straight 20-80 μm CNT |
| C | Muntz Brass | Inconel 718 | 0.1%Fe$_2$O$_3$ 0.1%Ni | 4h | 0.15 | 96% curled CNT |
| D | Muntz Brass | Nichrome C | 0.1%Fe$_2$O$_3$ | 15h | 0.08 | 70% 10-30 μm CNT |
| **E** | Muntz Brass | Inconel 625 3 layers Inconel 600 | 0.1%Fe$_2$O$_3$ | 15h | 0.08 | 97% 20-50 μm CNT |
| F | Muntz Brass | Inconel 718 2 layers Inconel 600 | 0.1%Fe$_2$O$_3$ | 4h | 0.15 | 98% straight 100-500 μm CNT |
| G | Muntz Brass | Inconel 718 3 layers Inconel 600 | 0.1%Fe$_2$O$_3$ 0.1%Ni | 15h | 0.08 | 90% Curled CNT or fibers |
| H | Muntz Brass | Nichrome C | 0.1%Fe$_2$O$_3$ | 1h | 0.4 | 96% Straight 100-200 μm CNT |
| I | Monel | Nichrome C | 0.1%Fe$_2$O$_3$ | 1h | 0.4 | 97% Straight 20-50 μm CNT |
| J | Monel | Nickel | / | 2h | 0.2 | 70% thin 10-20 μm CNT Rest: Onions |
| K | Monel | Nichrome C | 0.1%Fe$_2$O$_3$ | 2h | 0.1 | 97% 30-60 μm straight CNT |
| L | Monel | Nichrome C | 0.5%Fe$_2$O$_3$ | 15h | 0.08 | ~25% curled CNT ~70% straight CNT |
| M | Monel | Iridium | 0.81% Cr | 18h | 0.08 | 97% thin 50-100 μm CNT |

At a low current density of 0.08 A/cm², with an electrolyte additive of 0.1 wt% Fe$_2$O$_3$, the conventional Muntz Brass and Nichrome electrodes exhibit a significant drop in CNT product purity to 70%. Coulombic efficiency tends to drop off with current density, and in this case the coulombic efficiency of the synthesis was 82%. Product purity can be increased by refining the mix of transition metals available during the electrolytes or increasing surface area. Alloy composition of the metals used as electrodes is presented in Table 2. Metal variation was further refined by combining the metals in Table 2 as anodes, for example using a solid sheet of one Inconel alloy, layered with a screen or screens of another Inconel alloy. This approach is utilized in the lowest row of Figure 2 (panels #E), which utilizes an anode of Inconel 625 with 3 layers of (spot welded) 100 mesh Inconel 600 screen, a return to a single electrolyte additive (0.1 wt % Fe$_2$O$_3$) and a very low current density of 0.08 A/cm². As seen in panels #E of the figure, the product is high purity (97%) and consists of 20-50 μm length CNTs, and the coulombic efficiency was 75%. Not shown, but included in Table 1 (Electrolysis #G), is that under the same electrode, and the same 0.08 A/cm² electrolysis conditions. However, with the electrolyte addition of both 0.1 wt% Fe$_2$O$_3$ and 0.1 wt% Ni at J= 0.15 A/cm², the product is twisted CNTs as in Figure 2 panels #C, the purity is 96%, and the coulombic efficiency is 80%.

**Table 2.** Compositions of various alloys used (weight percentage).

| Alloy | Ni % | Fe% | Cu% | Zn% | Cr % | Mo% | Nb & Ta % |
|---|---|---|---|---|---|---|---|
| Nichrome C | 60 | 24 | | | 16 | | |
| Nichrome A | 80 | | | | 20 | | |
| Inconel 600 | 52.5 | 18.5 | | | 19.0 | 3.0 | 3.6 |
| Inconel 718 | 72% min | 6-10 | | | 14-17 | | |
| Inconel 625 | 58 | 5 max | | | 20-23 | 8-10 | 4.15-3.15 |
| Monel | 67 | | 31.5 | | | | |
| Muntz Brass | | | 60 | 40 | | | |

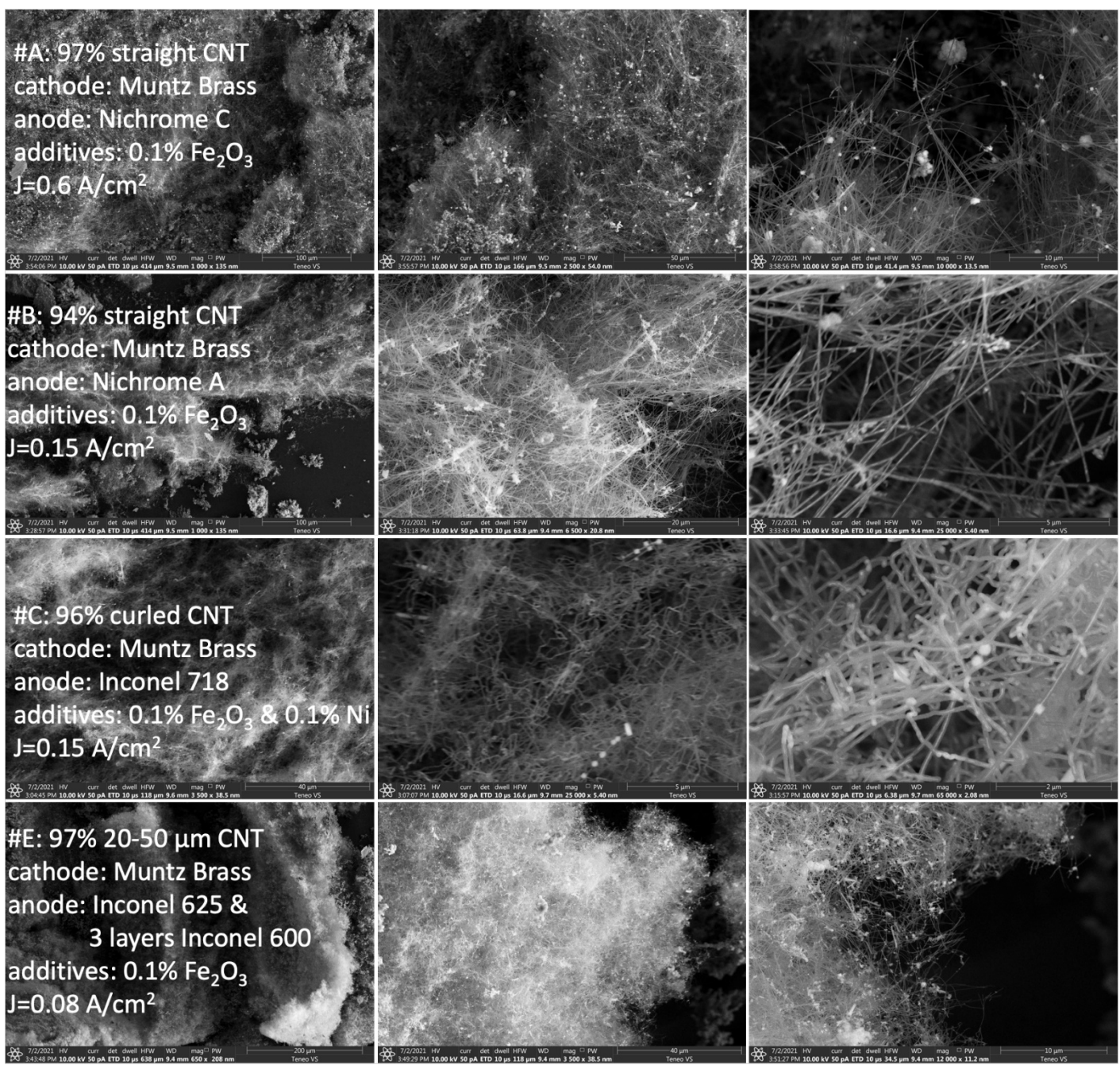

**Figure 2.** SEM of the synthesis product of high purity, high yield carbon nanotubes under a variety of electrochemical conditions by electrolytic splitting of $CO_2$ in 770°C $Li_2CO_3$. The washed product is collected from the cathode subsequent to the electrolysis described in Table 1. Moving left to right in the panels, the product is analyzed by SEM with increasing magnification. Scale bars in panels (starting from left) are for panels A: 100, 50 and 10 μm; for panels B: 100, 20 and 5 μm; for panels C: 40, 5 and 2 μm; for panels E: 200, 40 and 10 μm.

*2.2. Electrolytic conditions to synthesize high purity, high yield CNTs from $CO_2$.*

The syntheses listed in Table 1 delineate the electrochemical growth conditions for the high purity growth of carbon nanotubes each exhibiting the characteristic concentric multiple graphene cylindrical walls. This is observed in Figure 3, which presents Transmission Electron Microscopy (TEM) and High Angle Annular Dark-Field TEM (HAADF) of a typical example (the product of Electrolysis #E as further described in Table 1 and Figure 2), and which provides general structural and mechanistic information of carbon nanotubes synthesized by molten electrolysis. As seen in the top row of the figure, the carbon nanotubes are formed by successive, concentric layers of cylindrical graphene. The graphene is identified by its characteristic inter-graphene layer separation of 0.33 to 0.34 nm as measured in the figure by the spacing between the dark layers of uniform blocked electron transmission on the magnified top right side of the figure. This CNT has an outer diameter of 74 nm, and inner diameter of 46 nm and by counting dark rows it is determined that the number of graphene layers in this CNT is 41. The right side of the third row of the figure measures the carbon elemental profile of the CNT. This profile is swept laterally from the tube's exterior (no carbon) through the left wall (carbon), then through the void of interior of the tube (low carbon from the exterior backside wall), then through the right wall (carbon) and finally to the exterior of the tube on the outer left side (no carbon). Also the integrated elemental profile of area 1 of this panel is shown, which exhibits 100.0% carbon (fit error 1.3%).

In Figure 3 on the right side of row 2, the parallel 0.34 nm spacing for the graphene layers in the CNT walls is again observed. This panel also includes dark areas of metal trapped within the CNT, and which serves as a snapshot in time of the growth of the CNT. In the third row of the figure HAADF analysis of Area #1 has an elemental composition for this area including the walls with the trapped interior metal of 94.4% carbon, 2.5% Fe and 3.2% Ni as distributed according to the individual C, Fe and Ni HAADF maps included in the figure. The second row of the figure also shows the tip of the CNT, which includes trapped metal. The transition metal serves as a nucleating agent, which supports formation of the curved graphene layers shown at the tip of the CNT, which is a major component of the CNT growth mechanism. While occurring in an entirely different physical chemical environment than chemical vapor deposition (CVD), this molten carbonate electrolysis process of transition metal nucleated growth CNTs appears to be similar to those noted to occur for CVD CNT growth. This is despite the fact that CVD is a chemical/rather than electrochemical process, and occurs at the gas/solid, rather than liquid/solid interface.

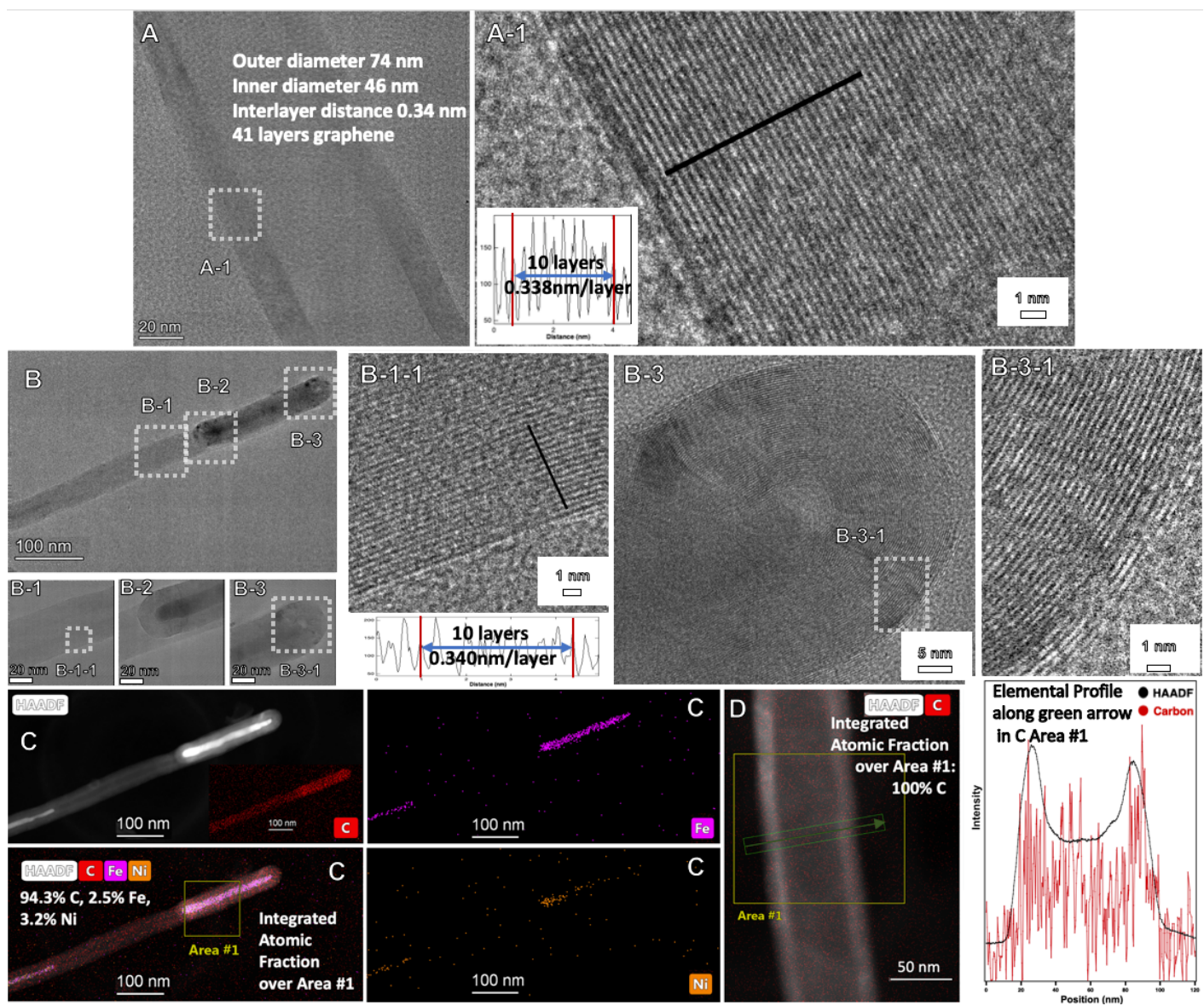

**Figure 3.** TEM and HAADF of the synthesis product of high purity, high yield carbon nanotubes under the Electrolysis #E (Table 1) electrochemical conditions by electrolytic splitting of $CO_2$ in 770°C $Li_2CO_3$. In the top row the product is analyzed by TEM with scale bars of 20 nm (left panel) or 1 nm (right). Moving left to right in the second row there are scale bars of 100, 5, 5 and 1 nm. Third row's scale bars are 100 or 50 nm. Bottom row scale bars are 20, 1 and 1 nm. Panels: A , B, B-1, B-2, B-3, B-3-1 TEM; A-1, B-1-1 TEM with measured graphene layer thickness. C: Elemental HAADF elemental analysis, D: HAADF element analysis with (right side) elemental profile).

*2.3 Tailored electrochemical growth conditions producing high aspect ratio CNTs from $CO_2$.*

Figure 4 presents the product's SEM of the electrochemical configuration which yields the longest (100 to 500 μm long) and highest purity (98%) CNTs at high coulombic efficiency (99.5%) of those studied here (described as Electrolysis #F in Table 1). As with the previous configuration that yielded nearly as high purity, but shorter, CNTs. The synthesis used an 0.1 wt% $Fe_2O_3$ additive to the $Li_2CO_3$ electrolyte, a Muntz Brass cathode and an Inconel 718 anode with layered Inconel 600 screen. However, this synthesis found an optimization in CNT purity and length using 2, rather than 3 layers of Inconel 600, and using a higher current density (0.4, rather than 0.08, A/cm²) and shorter electrolysis time (4, rather than 15, hours). With a diameter of < 0.2 μm, these CNTs can have an aspect ratio of > 1,000. As correlated with the alloy composition in Table 1, the smaller number of Inconel 600 layers reflects the need for inclusion of anodic molybdenum available in that alloy, but at a controlled, lower concentration, to achieve the resultant high purity, high aspect ratio CNTs. As seen in the figure, the CNTs are densely packed and largely parallel, and as discussed in sect. 2.5 would comprise a useful candidate for use in nano-filtration.

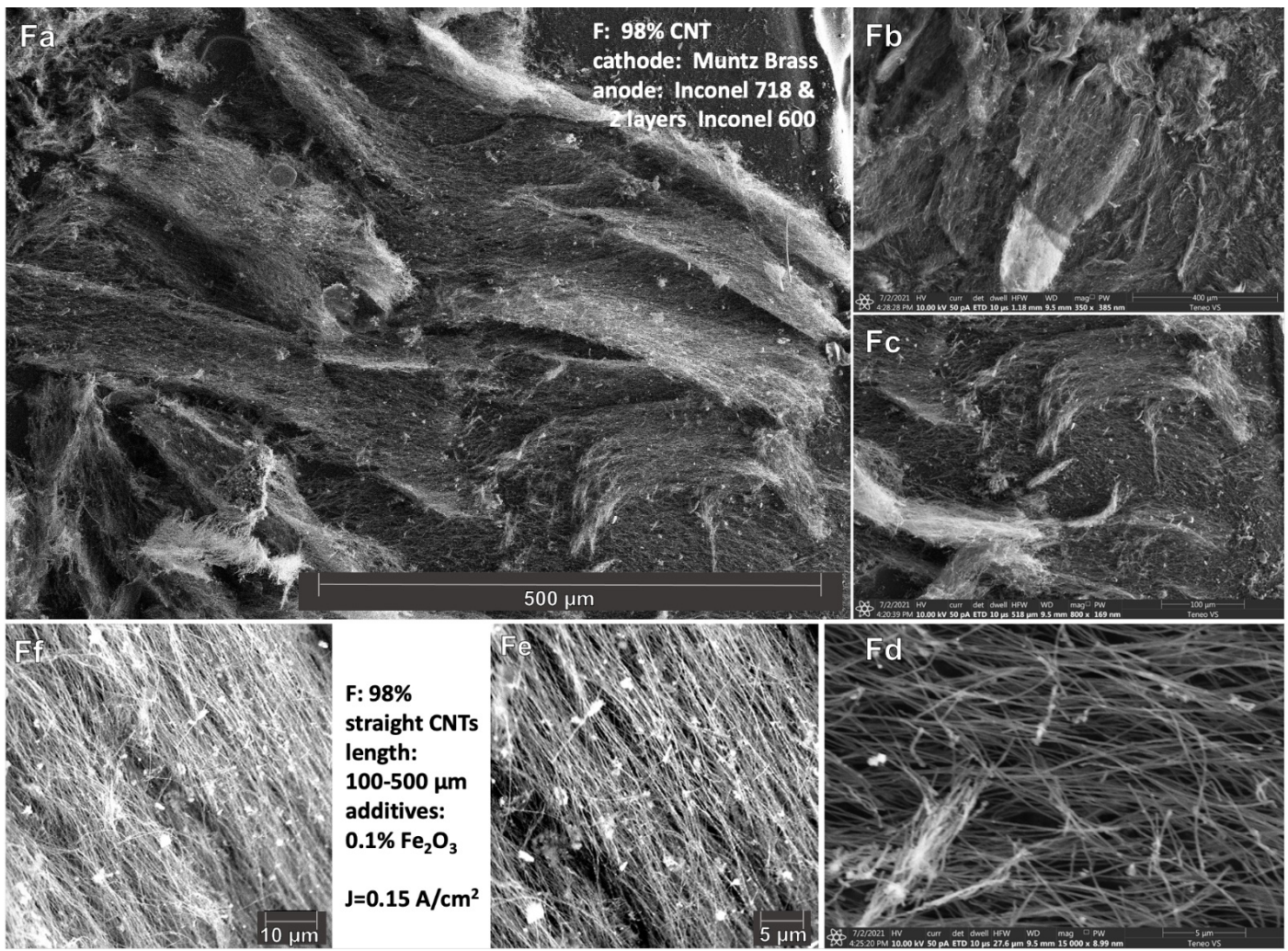

**Figure 4.** SEM of the synthesis product of high aspect ratio (and high purity and yield) carbon nanotubes prepared by electrolysis #F in Table 1, splitting $CO_2$ in 770°C $Li_2CO_3$. Moving left to right in the panels, the product is analyzed by SEM with increasing magnification. Scale bars in panels Fa-Ff (clockwise from top) are 500, 400, 100, 5, 5 and 10 μm.

Figure 5 presents TEM and HAADF probes of the high aspect ratio CNT product of Electrolysis #F (as described in Table 1, and by SEM in Figure 4). As seen on the right side of the middle row of the figure, the CNT walls consist of parallel carbon (layers separated by the characteristic 0.33 – 0.24 nm graphene layer spacing. As seen in the elemental analysis of Area #1 in the lowest row, areas consist of hollow tubes composed of 100% carbon. However, as seen in the TEM of the top two rows and in the bottom row as the HAADF elemental profiles, there are also extensive portions of the tubes that are intermittently filled with metal. In the bottom row of the figure, a lateral cross sectional elemental CNT profile scanned through Area #2 from the outside, through the CNT and then out the opposite wall shows the wall is composed of carbon, while the inner region also contains iron as the dominant metal coexisting with some nickel.

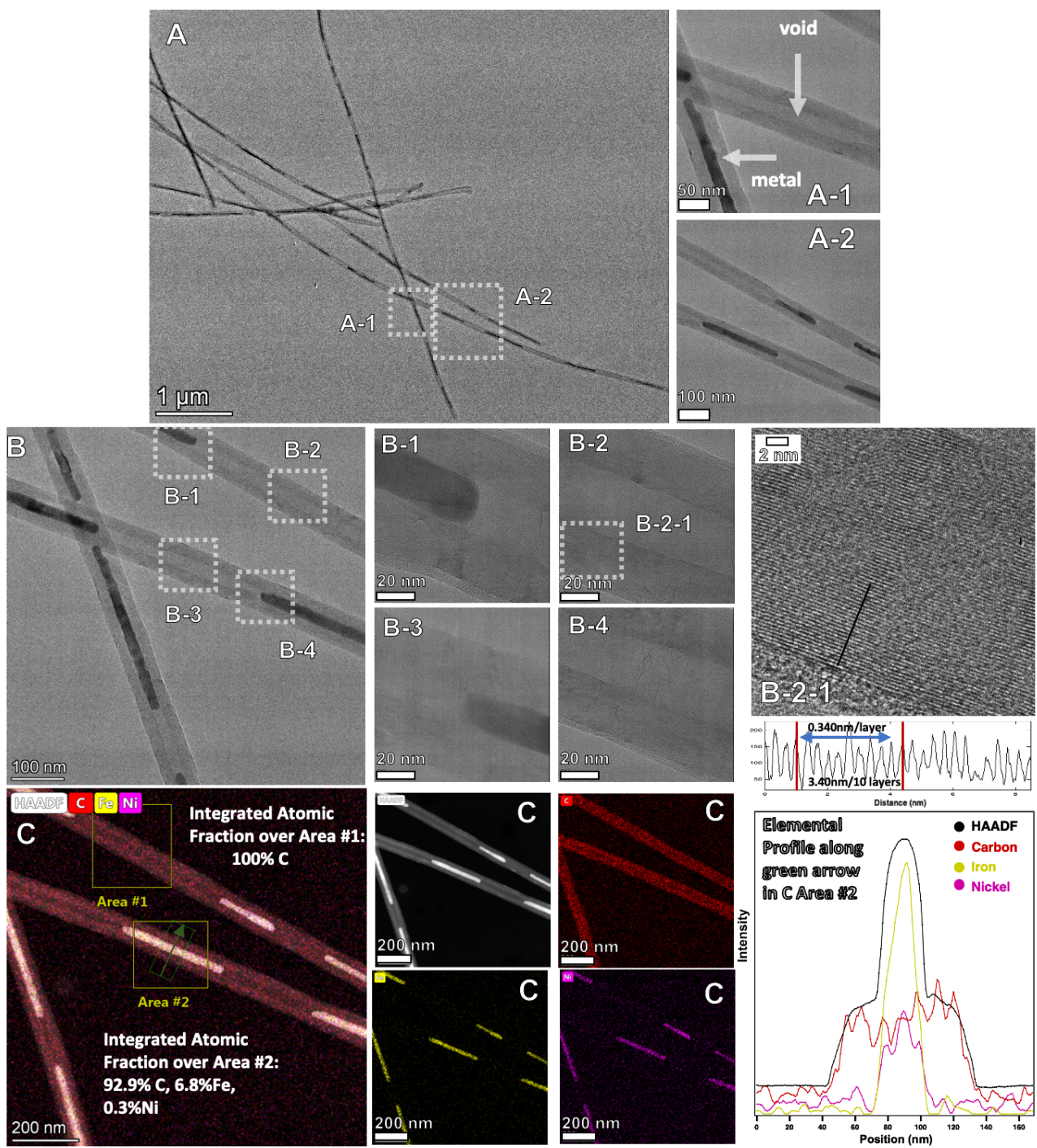

**Figure 5.** TEM and HAADF of the synthesis product of high purity, high yield carbon nanotubes under the Electrolysis #F (Table 1) electrochemical conditions by electrolytic splitting of $CO_2$ in 770°C $Li_2CO_3$. In the top row the product is analyzed by TEM with scale bars of 1 μm (left panel)

or 100 nm (right). Scale bars in the middle right moving left to right have scale bars of 50, 20 and 1 nm. HAADF measurements in the bottom panel each have scale bars of 200 nm. Panels: A, A-1, A-2, B, B-1, B-2, B-3, B-4:   TEM; B--2-1-1 TEM with measured graphene layer thickness. C: Elemental HAADF elemental analysis with (right side) elemental profile.

*2.4 Tailored electrochemical growth conditions producing thinner CNTs from $CO_2$.*

Figure 6 demonstrates additional electrochemical conditions which yield high purity carbon nanotubes by $CO_2$ molten electrolysis. In the top row, panels #H, as with high current density (Figure 2 panel #A), a moderate current density of 0.4 A/cm$^2$ (with the same electrolyte, a Muntz Brass cathode and a Nichrome C anode, yields high purity (96%) CNTs, that are somewhat longer (100-200 μm) at a coulombic efficiency approaching 100%. Switching the cathode material to Monel in the second row (Figure 6, panels a) yields shorter 20-50 μm CNTs with 97% purity and coulombic efficiency again approaching 100%. Not shown in the figure, but included in Table 1 (Electrolysis #D), is that a switch from Nichrome C to a pure nickel anode (while retaining the Monel Cathode, and with electrolyte additives at J=0.2 A/cm$^2$) leads to a substantial drop in CNT purity to 70% with the remainder of the product consisting of nano-onions. A drop of current density from 0.4 A to 0.1 A/cm$^2$ in Figure 6 panels #K yields 97% purity CNTs of length 30-60 μm with only a small drop of coulombic efficiency to 97%. In a single panel of #L located in the lower left corner of Figure 6, an overabundance of $Fe_2O_3$ is added which has previously been observed to lose control of the synthesis specificity [45]. In this case, the total purity of CNTs remains high at ~95%, but this consists of two distinct morphologies of CNT in the product. The majority product at ~75% is twisted CNTs, and the minority product at ~20% is straight CNTs. Finally, in panels #M on the middle and right lowest row of Figure 6, a noble metal, iridium, is used as the anode (along with the Monel Cathode) at a low 0.08 A/cm$^2$ current density. Transition metals released from the anode, during its formation of a stable oxide over layer, can contribute to the transition metals ions that are reduced at the cathode and serve as nucleation points for the CNTs. This is not the case here due to high stability of the iridium. Instead as a single, high concentration transition metal, 0.81 wt% Cr, is made as the electrolyte additive. The product is highly pure (97%) CNTs that are the thinnest shown (< 50 nm diameter), are 50-100 μm long for an aspect ratio > 1,000, and formed at a coulombic efficiency of 80%.

SEM of several of the synthesis products, specifically Electrolyses #H, #B and #C, exhibit evidence of nodules that appear as "buds" attached to the CNTs. These buds are the most consistent in Electrolysis #H and are explored by TEM and HAADF in Figure 7. It is fascinating, as seen in the top row of the figure, that the buds generally have a spherical symmetry, and while not prevalent in the structure appear in a structure comparable to grape bunches growing on a vine. The buds generally contain a low level of the transition metal nucleating metal, such as the 0.3% Fe evident, and the rest of the structure is generally pure carbon, with an occasional metal core. This low level of metal used is easily removed by an acid wash. Previously introduced higher levels of Ni or Fe can lead to magnetic carbon nanotube with useful properties in recyclability, filtration, and shape-shifting materials among other applications [41]. As seen in the left side of the second row, the carbon nanotube walls continue to exhibit the regular 0.33 to 0.34 graphene interwall separation, as seen on the right side of the row, joining adjacent CNTs may have merged or distinct graphene structures. Similarly, as seen in the third row of Figure 2, adjacent buds on CNTs can have graphene walls which bend to join, and are shared, or as seen in the fourth row appear instead to be distinct (intertwined, not merged) structures.

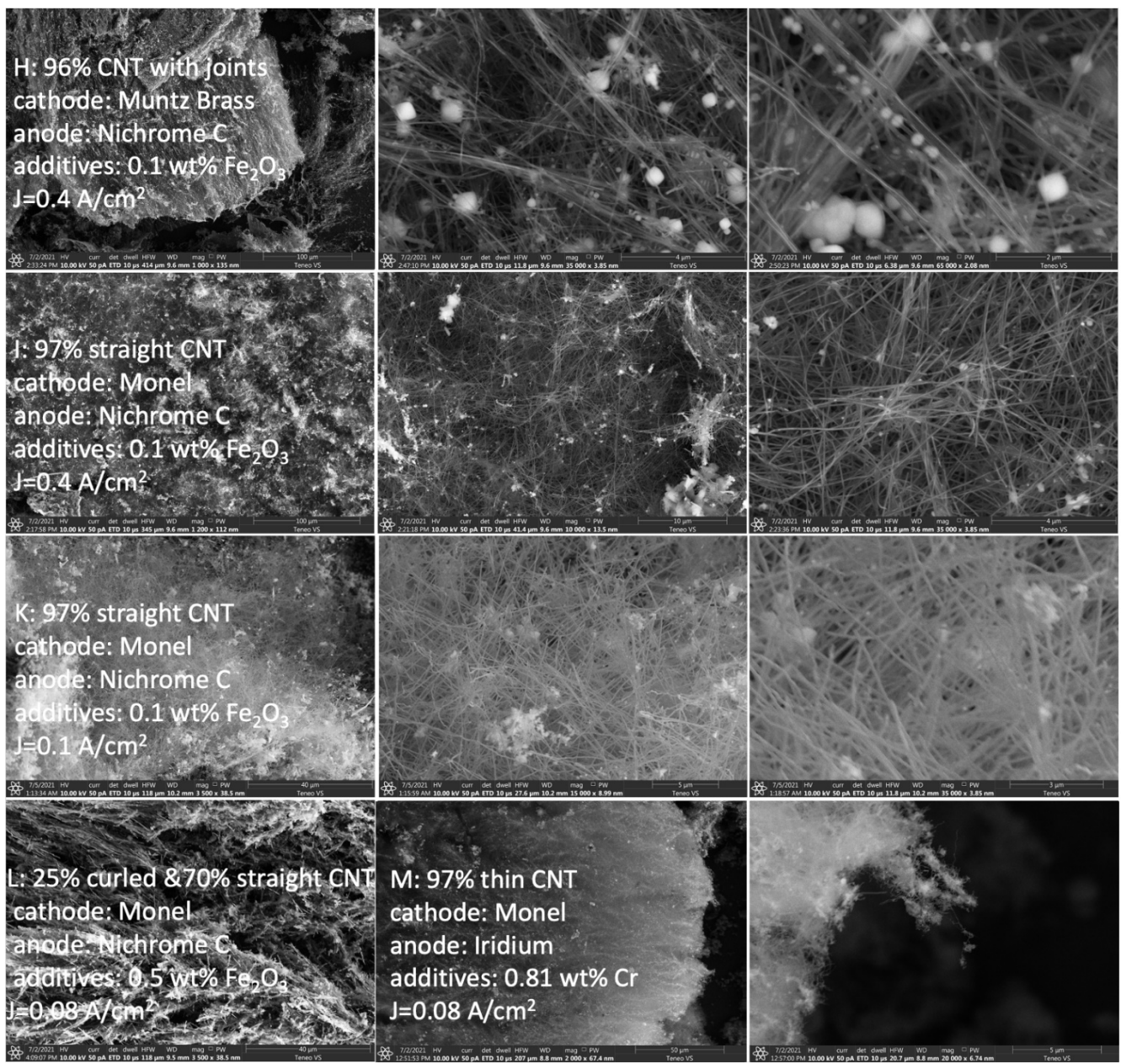

**Figure 6.** SEM of the synthesis product of high purity, high yield carbon nanotubes under a variety of electrochemical conditions by electrolytic splitting of $CO_2$ in 770°C $Li_2CO_3$. The washed product is collected from the cathode subsequent to the electrolysis described in Table 1. Scale bars (starting from left) are for panels J: 100, 4 and 2 μm; for panels I: 100, 10 and 4 μm; for panels K: 40, 5 and 3 μm; for panels L: 40, 50 and 5 μm.

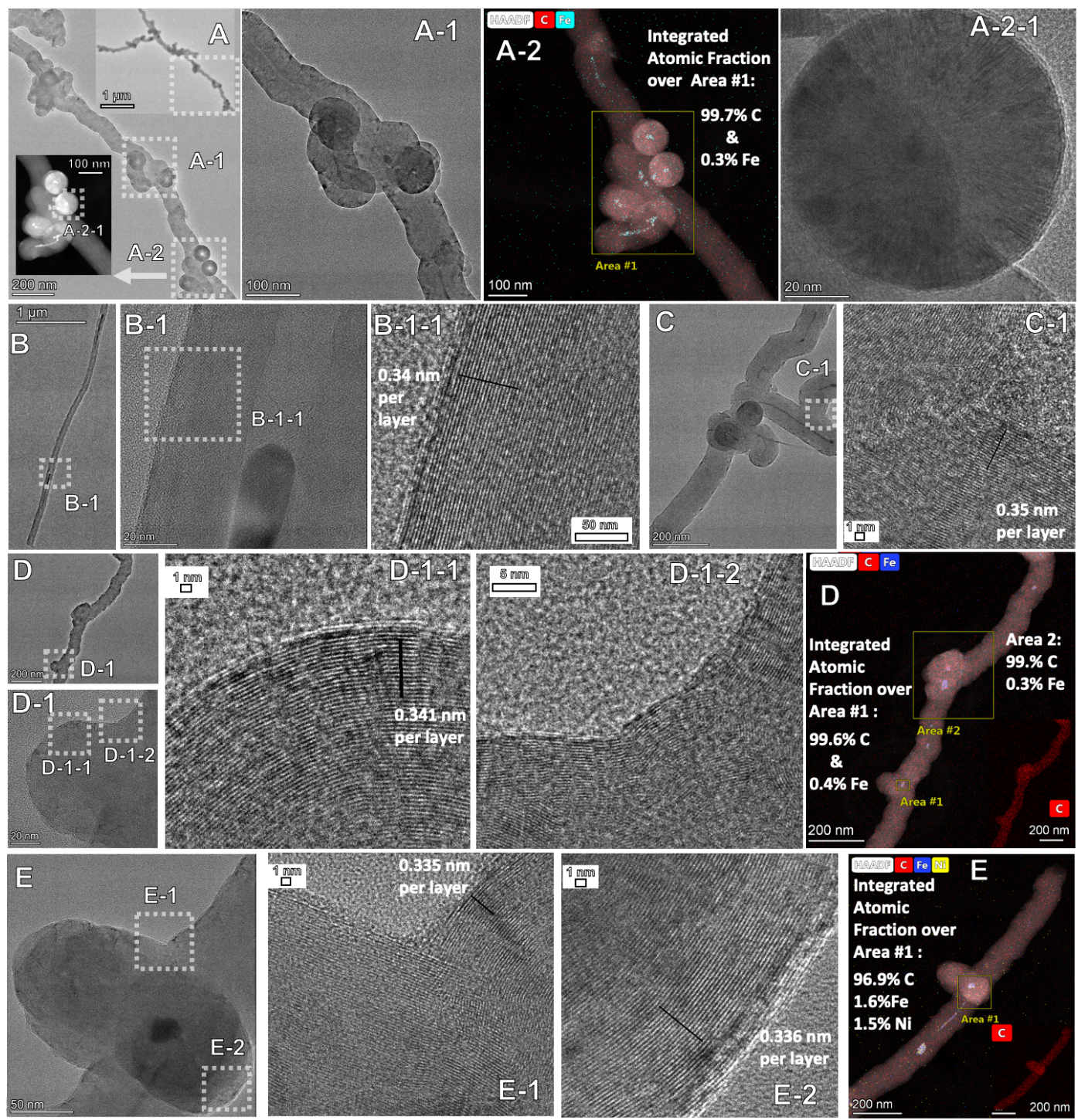

**Figure 7.** TEM and HAADF of the synthesis product of carbon nanotubes which exhibit nodules or buds under the Electrolysis #H (Table 1 and SEM on top row of Figure 6) electrochemical conditions by electrolytic splitting of $CO_2$ in 770°C $Li_2CO_3$. In the top row the product is analyzed with scale bars from left to right of 200, 100, 20 and 100 nm. Scale bars in the second right have scale bars of 1μm. then 20, 5, 200 and 1 nm. Third row scale bars are 200, 20, 1, 5 and 200 nm. Bottom row scale bars are 50, 1, 1 and 200 nm. Panels: A, A-1, A-2-1, B, B-1, B—1-1, C, C-1, C, C-1, D-1-1, D-1-2, E, E-1, E2: TEM; A-2, D, E: Elemental HAADF elemental analysis.

*2.5. Electrochemical conditions to synthesize macroscopic assemblies of CNTs.*

   In addition to individual CNTs, the final series of electrolyses generate useful macroscopic assemblies of CNTs. There has been interest in densely packed CNTs for nano-filtration, and also due to their high density of conductive wires as an artificial neural net [53-60,61-62]. CNTs aerogels have been reported as formed by CVD and/or also reported as formed within molds. Their sorbent properties have been investigated for applications such as cleanup of chemical leakage under harsh conditions. Those studies noted that such aerogel matrices, consisting of highly porous, intermingled CNTs, can be repeatedly compressed to a small fraction of their initial volume without damaging the structure of the carbon nanomaterials [52-56].

The term "Nanofiltration" was proposed in 1984 to solve the terminology problem for a selective reverse osmosis process that allows ionic solutes in a feed water to permeate through a membrane [57]. In addition to low energy consumption, with respect to alternative unit operations such as distillation and evaporation, thermal damage of heat-sensitive molecules can be minimized during the separation, due to the potential for low operating temperatures of nanofiltration [58]. Small diameter CNTs have been demonstrated for the ability to use as a nanofiltration or molecular sieve to selectively remove larger size molecules from smaller size molecules, such as the selective removal of cyclohexane from n-hexane [57]. There is a need to control the fabrication of macroscopic assemblies of CNTs to optimize their capabilities for nanofiltration, and CNT assemblies synthesized by CVD and laser ablation have been investigated 52-56,58-61].

An artificial neural network is a collection of interconnected nodes loosely model the neurons in a biological brain. Estimates of biologic neuronal density (rat) are in the range of 100 in a 100 µm cube ($10^5$/mm$^3$) on each side. Fabrication of an artificial neural network with structure that mimics that number of nanowires and nodes presents a challenge. However, this is in the same size domain of macroscopic assemblies of CNTs. For example, Gabay and co-workers have explored the engineered self-organization of neural networks using carbon nanotube clusters [38], emphasizing the need for improved pathways to fabricate and control macroscopic assemblies of CNTs.

The macroscopic assemblies observed in this study are referred to as nano-sponge, dense packed parallel CNTs, and nano-web CNTs in Table 3 and Figure 8. The Nano-sponge assembly is formed by Electrolyses #N with Nichrome C serving as both the cathode and the anode, with 0.81% Ni powder added to the 770°C $Li_2CO_3$ electrolyte, the initial current ramped upwards (5 min each at 0.008, 0.016, 0.033 and 0.067 A/cm$^2$), then a 4h current density of 0.2 A/cm$^2$ generating a 97% purity nanosponge at 99% coulombic efficiency. As previously described, long densely packed, parallel carbon nanotubes are produced in Electrolysis #F with a 0.1 wt% $Fe_2O_3$ additive to the $Li_2CO_3$ electrolyte, a Muntz Brass cathode and an Inconel 718 anode and 2 layers of Inconel 600 screen at 0.15 A/cm$^2$.

**Table 3.** Systematic variation of $CO_2$ splitting conditions in 770°C $Li_2CO_3$ to optimize formation of macroscopic assemblies of nanocarbons with densely packed carbon nanotubes.

| Electrolysis # | Cathode | Anode | Additives (wt% powder) | Electrolysis time | Current density (A/cm$^2$) | Product Description |
|---|---|---|---|---|---|---|
| N | Nichrome C | Nichrome C | 0.81% Ni | 4h | 0.2 | 97% nano-sponge CNT |
| F | Muntz Brass | Inconel 718 2 layers Inconel 600 | 0.1% $Fe_2O_3$ | 4h | 0.15 | 98% dense packed straight 100-500µm CNT |
| P | Muntz Brass | Nichrome C 3 layers Inconel 600 | 0.1% $Fe_2O_3$ | 15h | 0.08 | 97% 50-100µm nano-web CNT |
| Q | Monel | Nichrome C | 0.81% Ni | 3h | 0.2 | 92% 5-30µm nano-web CNT Rest: onions |

As opposed to the parallel assembly produced in Electrolysis #F, nano-web aptly describes the interwoven carbon nanotubes from Electrolyses #P and #Q, presented in the lower rows of Table 3 and Figure 8. Two different routes to the nano-web assembly are summarized. The first uses an 0.1% $Fe_2O_3$ additive, a Muntz Brass cathode and an Inconel 718 anode with 3 layers of Inconel 600 screen, at 0.08 A/cm² generating a nano-web with a purity of 97% at a coulombic efficiency of 79%. The second pathway uses an 0.81% Ni powder additive, a Monel cathode and Nichrome C anode, at 0.28 A/cm² generating a nano-web with a purity of 92% at a coulombic efficiency of 93%.

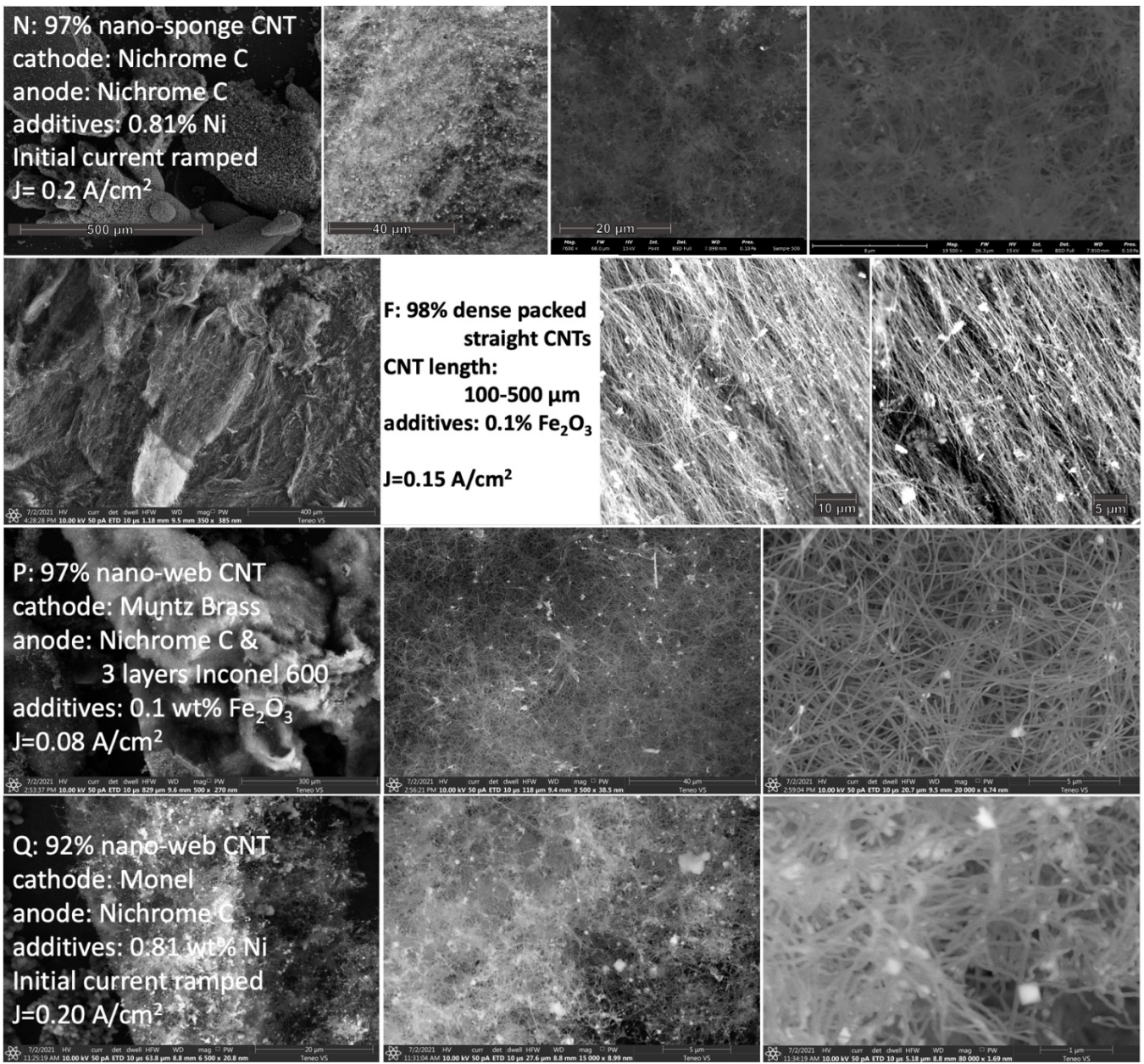

**Figure 8.** SEM of the synthesis product consisting of carbon nanotubes arranged in various packed macroscopic structures that are amenable to nano-filtration. The washed product is collected from the cathode subsequent to the electrolysis described in Table 3. Moving left to right in the panels, the product is analyzed by SEM with increasing magnification. Scale bars in panels These include nano-sponge, dense packed straight, and nano-web CNTs. Moving left to right in the panels, the product is analyzed by SEM with increasing magnification. Scale bars in panels (starting from left) are for panels 5: 500, 40, and 20 and 8 μm; for panels M: 400, 10 and 5 μm; for panels d: 300, 40 and 5 μm; for panels 5: 500 μm 40 20 and 8 μm.

The dense packed straight CNTs have an inter-CNT spacing ranging from 50 to 300 nm, moreover the CNTs are highly aligned providing unusual nanofiltration opportunites for both this size domain, and for an opportunity to filter 1D from 3D morphologies. The nano-sponge do not have this alignment feature, and from Figure 8, provide nanofiltration pore size of 100 to 500 nm, while the nano-web product provides nanofiltration with pore size of 200 nm to 1 μm. Future studies will investigate the effectiveness of this portfolio of macroscopic CNT assemblies for nanofiltration.

*2.6. Raman and XRD characterization of the CNTs and their macro-assemblies.*

Figure 9 presents the Raman spectra effect of variation of the CNT electrolysis conditions on the CNT assembly products from $CO_2$ electrolysis in 770°C $Li_2CO_3$. The Raman spectrum exhibits two sharp peaks ~1350 and ~1580 cm$^{-1}$, which correspond to the disorder-induced mode (D band) and the high frequency $E_{2G}$ first order mode (G band), respectively and an additional peak, the 2D band, at 2700 cm$^{-1}$. In the spectra, the graphitic fingerprints lie in the 1880-2300 cm$^{-1}$ and are related to different collective vibrations of sp-hybridized C-C bonds.

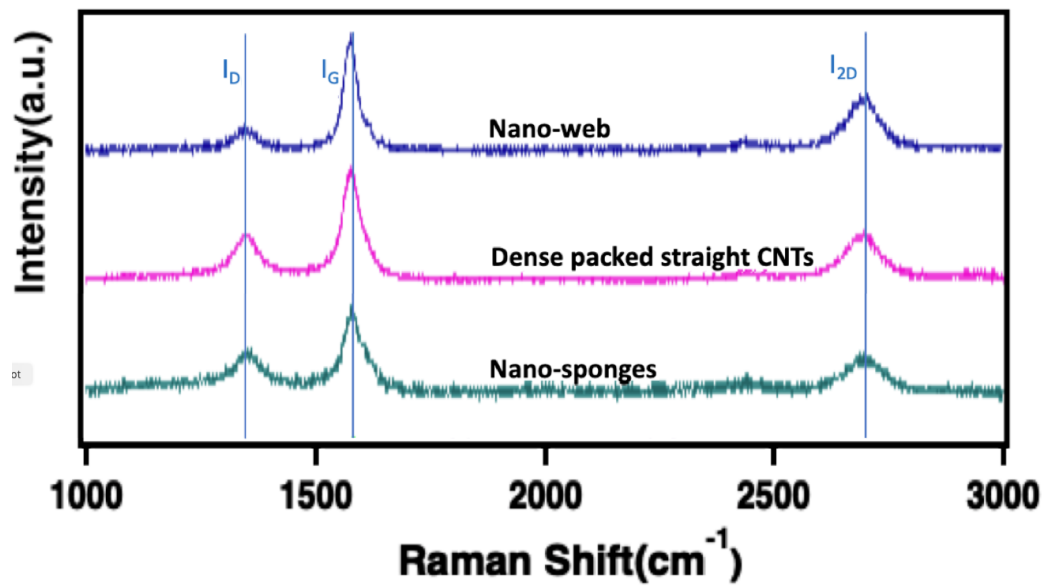

**Figure 9.** Raman of the synthesis product consisting of various labeled carbon nanotube assemblies synthesized by the electrolytic splitting of $CO_2$ in 770°C $Li_2CO_3$ with a variety of systematically varied electrochemical conditions described in Table 4.

Interpretation of the Raman spectra provides insight into potential applications of the various carbon allotropes. From Figure 9, the intensity ratio between D band and G band ($I_D/I_G$) is calculated, or an observed shift in $I_G$ frequency, are useful parameters to evaluate the relative number of defects and degree of graphitization are presented in Table 4. Note in particular, that of the nano-sponge, nano-web and dense packed CNT assemblies described in Figure 8 and Table 3, that the nano-web CNT assembly exhibits low disorder with $I_D/I_G$ = 0.36 in Table 4, the dense packed CNT assembly exhibits intermediate disorder with $I_D/I_G$ = 0.49, and the nano-sponge exhibits the highest disorder with $I_D/I_G$ = 0.62 and while accompanied by a shift in $I_G$ frequency.

That is for the assemblies with increasing $I_D/I_G$ ratio:

CNT nano-web  <  Dense packed CNT  <  CNT nano-sponge

Previously, increased concentrations of iron oxide added to the $Li_2CO_3$ electrolyte had correlated with an increasing degree of disorder in the graphitic structure [38]. It should be noted that these defect levels each remain relatively low as the literature is replete with reports of multiwalled carbon nanotubes made by other synthetic processes with $I_D/I_G > 1$. Lower defects are associated with for applications which require high electrical and strength, while high defects are associated with for applications which permit high diffusivity through the structure such as those associated with increased intercalation and higher anodic capacity in Li-ion batteries and higher charge supercapacitors.

Table 4. Raman spectra of a diverse range of carbon allotropes and macro-assemblies formed by molten electrolysis.

| CO$_2$ Molten Electrolysis Product Description | $\nu_D$(cm$^{-1}$) | $\nu_G$(cm$^{-1}$) | $\nu_{2D}$(cm$^{-1}$) | $I_D/I_G$ | $I_{2D}/I_G$ |
|---|---|---|---|---|---|
| Nano-web | 1342.5 | 1577 | 2689.6 | 0.28 | 0.50 |
| Dense packed CNTs | 1342.5 | 1577.4 | 2694.8 | 0.46 | 0.49 |
| Nano-sponge | 1352.5 | 1580.6 | 2687.3 | 0.67 | 0.62 |

Along with the XRD library of relevant compound spectra, XRD is presented in Figure 10 of the CNT assembly products, prepared as described in Figure 8 and Table 3. Each of the spectra exhibit the strong diffraction peak at $2\theta = 27°$ characteristic of graphitic structures. The nano-sponge XRD spectra is distinct from the others having a dominant peak at $2\theta = 43°$, indicating the presence of iron as $Li_2Ni_8O_{10}$ and chromium as $LiCrO_2$ by XRD spectra match. XRD of this nano-sponge exhibits little or no iron carbide. On the other hand, both the nano-web and dense packed straight CNTs exhibit additional significant peaks at $2\theta = 42$ and $44°$ indicative of the presence of iron carbide, $Fe_3C$. The diminished presence of defects previously noted by the Raman spectra for the other dense packed CNTs along with the XRD presence of $Li_2Ni_8O_{10}$, $LiCrO_2$ and $Fe_3C$ provide evidence that the co-presence of Ni, Cr and Fe as nucleating agents can diminish defects in the CNT structure compared to Ni and Cr alone.

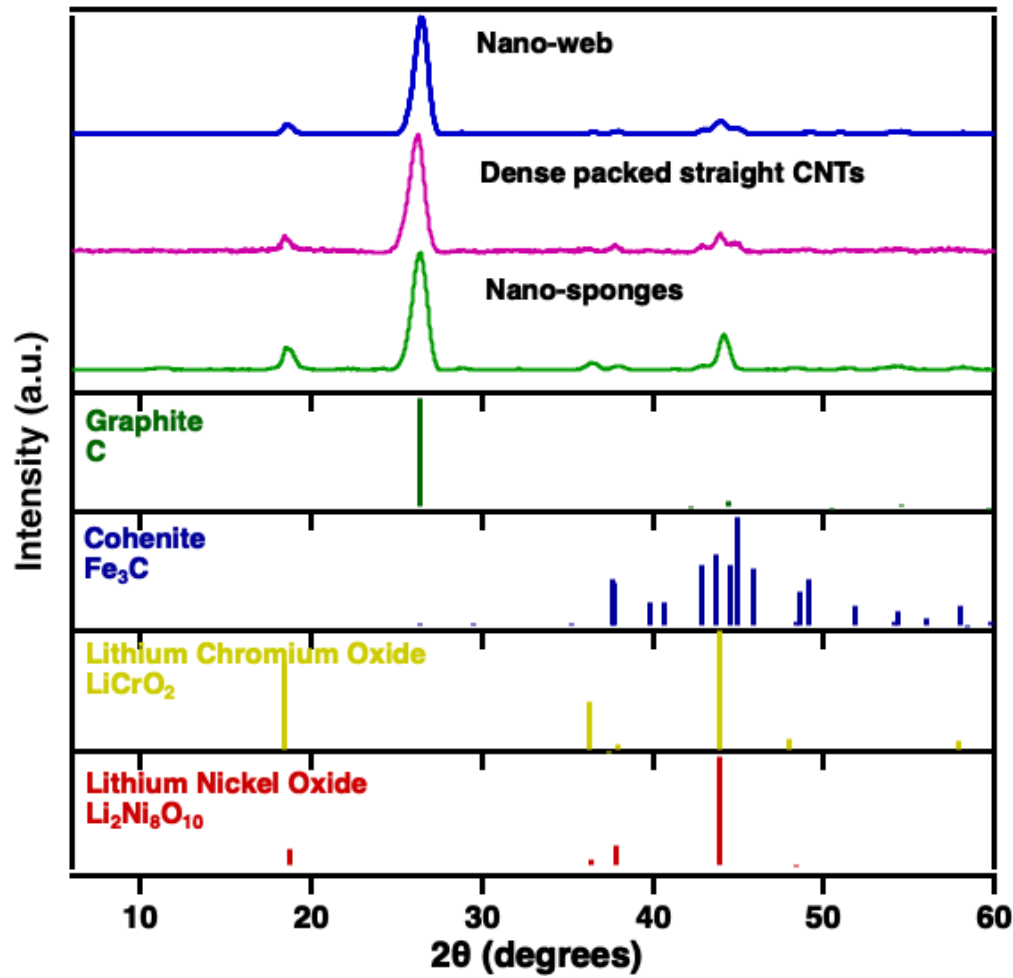

**Figure 10.** XRD of the synthesis product consisting of various labeled carbon nanotube assemblies synthesized by the electrolytic splitting of $CO_2$ in 770°C $Li_2CO_3$ with a variety of systematically varied electrochemical conditions described in Table 4.

## 3. Materials and Methods

*3.1. Materials.*

Lithium carbonate ($Li_2CO_3$, 99.5%), lithium oxide ($Li_2O$, 99.5%), lithium phosphate $Li_3PO_4$ ($Li_3PO_4$, 99.5%), iron oxide ($Fe_2O_3$, 99.9%, Alfa Aesar), and boric acid ($H_3BO_3$, Alfa Aesar 99+%) are used as electrolyte components in this study. For electrodes, Nichrome A (0.04-inch-thick), Nichrome C (0.04-inch-thick), Inconel 600 (0.25-in thick), Inconel 625 (0.25-in thick), Monel 400, Stainless Steel 304 (0.25-in thick), Muntz Brass (0.25-in thick), were purchased from onlinemetals.com. Ni powder is 3-7 µm (99.9%, Alfa Aesar). Cr powder is <10 µm (99.2%, Alfa Aesar). Co powder is 1.6 µm (99.8%, Alfa Aesar). Iron oxide is 99.9% $Fe_2O_3$ (Alfa Aesar). Co powder is 1.6 µm (99.8%, Alfa Aesar). Inconel 600 (100 mesh) was purchased from Cleveland Cloth. The electrolysis is a conducted in a high form crucible >99.6% alumina (Advalue).

*3.2. Electrolysis and purification.*

Specific electrolyte compositions of each electrolyte are described in the text. The electrolyte is pre-mixed by weight in the noted ratios then metal or metal oxide additives added if used. The cathode is mounted vertically across from the anode and immersed in the electrolyte. Generally, the electrodes are immersed subsequent to electrolyte melt. For several, noted, electrolyses, once melted, the electrolyte was maintained at 770°C ("aging" the electrolyte) prior to immersion of the electrolytes followed by immediate elec-trolysis. Generally, the electrolysis is driven with a described constant current density. As noted, for some electrolyses, the current density is ramped in several steps building to the applied electrolysis current, which is then maintained at a constant current density. Instead, most of the electrolyses are initiated, and held, at a single constant current. The electrolysis temperature is 770 °C using $CO_2$ directly from the air. In the C2CNT process, the electrolytic splitting can occur as direct air carbon capture without $CO_2$ pre-concentration [31-34, 36-44,47-51], or with concentrated $CO_2$, or $CO_2$ exhaust gas including during the scale-up of this process, in which the $CO_2$ transformation to CNTs was awarded the 2021 Carbon XPrize XFactor award for producing the most valuable product from $CO_2$ [32,45,46,62-64]. In this study, as the electrolysis cell directly captures $CO_2$ from the air via equation 1, no additional introduction of $CO_2$ is needed. A simple measure of sufficient $CO_2$ uptake is whether the electrolyte level falls during the course of the electrolyte. As mentioned, the $^{13}C$ isotope of $CO_2$ was previously used to track carbon through the C2CNT process from its origin ($CO_2$ as a gas phase reactant) through its transformation to a CNT or carbon nanofiber product, and that the $CO_2$ originating from the gas phase serves as the renewable C building blocks in the observed CNT product [32]. If $CO_2$ uptake is insufficient, then the carbonate electrolyte is instead consumed in accord equation 2, rather than renewed in accord with equations 1 and 2 in tandem. For example, when conducting at high electrolysis rates of 1 A / cm$^2$ or greater (not the situation of this study), then gas containing $CO_2$ must be bubbled into the electrolyte, otherwise electrolyte is consumed and the level of electrolyte visibly falls [40].

*3.3. Product characterization.*

The raw product is collected from the cathode after the experiment and cool down, followed by an aqueous wash procedure that removes electrolyte congealed with the product as the cathode cools. The washed carbon product is separated by vacuum filtration. The washed carbon product is dried overnight at 60 °C oven yielding a black powder product.

The coulombic efficiency of electrolysis is the percent of applied, constant current charge that was converted to carbon determined as:

100% x $C_{experimental}$ / $C_{theoretical}$ (4)

This is measured by the mass of washed carbon product removed from the cathode, $C_{experimental}$, and calculated from the theoretical mass, $C_{theoretical}$ = (Q/nF) x (12.01 g C mol$^{-1}$) which is determined from Q, the time integrated charged passed during the electrolysis, F, the Faraday (96485 As mol$^{-1}$ e$^-$), and the n = 4 e$^-$ mol$^{-1}$ reduction of tetravalent carbon consistent with equation 2.

*Characterization:* The carbon product was washed, and analyzed by PHENOM Pro Pro-X SEM (with EDX), FEI Teneo LV SEM, and by FEI Teneo Talos F200X TEM (with EDX). XRD powder diffraction analyses were conducted with a Rigaku D=Max 2200 XRD diffractometer and analyzed with the Jade software package. Raman spectra were collected with a LabRAM HR800 Raman microscope (HORIBA). This Raman spectrometer/microscope uses an incident laser light with a high resolution of 0.6 cm$^{-1}$ at 532.14 nm wavelength.

## 4. Conclusions

Molten carbonate electrolysis of $CO_2$ provides an effective path for the C2CNT synthesis of CNTs and macroscopic CNT assemblies. This study has explored a variety of electrochemical configurations, systematically varying electrode composition, electrode current density and electrolysis time, current ramping initiation, and variation of electrolyte additives and their concentrations. The highest observed CNT purity synthesis (97%) utilized a specialized anode consisting of 2 layers of high surface area inconel 600 (screen) on Inconel 718, a Muntz Brass cathode, with an 0.1 wt% $Fe_2CO_3$ additive to the 770°C $Li_2CO_3$ electrolyte, and with the electrolysis current conducted for 4 hours at an intermediate current density (without current ramping) of 0.15 mA/cm$^2$. The product, as analyzed by SEM, was aligned CNTs with a length of 100 to 500 μm and an aspect ratio of over 1000. The anode, cathode and electrolyte additive choice are an effective method for controlling the transition metal nucleation which is critical to high purity electrolytic CNT growth.

In a sister paper [65], slight variations of the synthesis parameters will form a variety of new high-purity, non-CNT nanocarbon allotropes. All syntheses in the study here produced a majority of CNTs, but the morphology of the CNT product changed widely with synthesis conditions. Depending on the synthesis conditions here alternate products of CNTs as short as 10 to 30 μm, and curled, rather than straight, or mixed with carbon nano-onions were observed. The high purity product exhibited a sharp XRD graphic peak, and a low Raman $I_D/I_G$ ratio indicative of low defects in the carbon structure. The XRD also contained iron carbide, and nickel and chromium lithium oxides, which from TEM are located within the CNT.

TEM HAADF shows that the inner core of CNT length is generally free of metals (void, with the walls 100% carbon), but in other areas the void is filled with transition metal. As produced by molten carbonate electrolysis, the CNT walls are conclusively shown to be comprised of highly uniform concentric, cylindrical graphene layers with a graphene characteristic, inter-layer spacing of 0.33 to 0.34 nm. When the internal transition metal is within the CNT tip, the layered CNT graphene walls are observed to bend in a highly spherical fashion around the metal supporting the transition metal nucleated CNT growth mechanism. Several syntheses had unusual nodules, many of them highly spherical, on the CNT, generally comprised of carbon and containing a low level of internal transition metal. Generally, intersecting CNTs did not merge, but in a few cases graphene layers bends to become part of the CNT intersection consistent with occasional, related growth of intersecting CNTs, such as branching, can occur.

The study also demonstrates new syntheses of assemblies of CNTs by the C2CNT process, with structural implications towards their potential applications for nano-filtration and neural nets and demonstrated pores sizes ranging from 50 nm to 1μm.

**Author Contributions:** Conceptualization, S.L. and G.L.; methodology, S.L., XL, G.L and X.W.; writing S.L. and G.L..

**Funding:** C2CNT LLC funded this research through the C2CNT LLC XPrize support funding.

**Data Availability Statement:** The authors confirm that the $data$ supporting the findings of this study are available within the article.

**Conflicts of Interest:** The authors declare no conflict of interest.